\documentclass[11pt,fleqn]{article}
\usepackage{amsfonts,amssymb,cite}
\usepackage{graphicx}

\usepackage{amsfonts,amssymb,cite}
\usepackage{epsf,graphicx}

\newcommand{\bls}[1]{\renewcommand{\baselinestretch}{#1}}

\def\noi{\noindent}
\def\jnumber#1#2{\thispagestyle{empty} \noi\unitlength=1mm
    	\begin{picture}(178,10)
            \put(177,15){\llap{\large Grav. Cosmol. \,#1 #2}}
                    \end{picture}}

\newcommand{\Title}[1]{\noi {{\Large\bf #1}}\\[1ex]}

\def\Aunames#1{\noi{\bf #1}}
\def\auth#1{${}^{#1}$}
\def\Addresses#1{\medskip\noi \protect
	\begin{description}\itemsep -3pt {\it #1} \end{description}}
\def\addr#1#2{\item[${}^{#1}$]{\it #2}}

\newcommand{\Abstract}[1]{\vskip 2mm \begin{center}
        \parbox{16.4cm}{\small\noi #1} \end{center}\medskip}

\newcommand{\foom}[1]{\protect\footnotemark[#1]}
\newcommand{\foox}[2]{\footnotetext[#1]{#2}\addtocounter{footnote}{1}}
\def\email#1#2{\footnotetext[#1]{e-mail: #2}\addtocounter{footnote}{1}}

\def\Talk{\foox 1 {Talk given at the International Conference RUDN-10,
	   June 28 --- July 3, 2010, PFUR, Moscow}}

\def\nq{\hspace*{-1em}}
\def\nqq{\hspace*{-2em}}
\def\nhq{\hspace*{-0.5em}}

\def\cm{\hspace*{1cm}}
\def\inch{\hspace*{1in}}

\def\Jl#1#2{#1 {\bf #2},\ }

\def\ApJ#1 {\Jl{Astroph. J.}{#1}}
\def\CQG#1 {\Jl{Class. Quantum Grav.}{#1}}
\def\DAN#1 {\Jl{Dokl. AN SSSR}{#1}}
\def\GC#1 {\Jl{Grav. Cosmol.}{#1}}
\def\GRG#1 {\Jl{Gen. Rel. Grav.}{#1}}
\def\JETF#1 {\Jl{Zh. Eksp. Teor. Fiz.}{#1}}
\def\JETP#1 {\Jl{Sov. Phys. JETP}{#1}}
\def\JHEP#1 {\Jl{JHEP}{#1}}
\def\JMP#1 {\Jl{J. Math. Phys.}{#1}}
\def\NPB#1 {\Jl{Nucl. Phys. B}{#1}}
\def\NP#1 {\Jl{Nucl. Phys.}{#1}}
\def\PLA#1 {\Jl{Phys. Lett. A}{#1}}
\def\PLB#1 {\Jl{Phys. Lett. B}{#1}}
\def\PRD#1 {\Jl{Phys. Rev. D}{#1}}
\def\PRL#1 {\Jl{Phys. Rev. Lett.}{#1}}

\def\al{&\nhq}
\def\lal{&&\nqq {}}
\def\eq{Eq.\,}

\def\beq{\begin{equation}}
\def\eeq{\end{equation}}
\def\bear{\begin{eqnarray}}
\def\bearr{\begin{eqnarray} \lal}
\def\ear{\end{eqnarray}}
\def\earn{\nonumber \end{eqnarray}}

\def\nnn{\nonumber\\ \lal }

\def\yy{\\[5pt] {}}

\def\eql{\al =\al}

\def\tst{\textstyle}

\def\fract#1#2{{\tst\frac{#1}{#2}}}

\def\half{{\fract{1}{2}}}

\def\d{\partial}

\def\diag{\mathop{\rm diag}\nolimits}

\def\const{{\rm const}}

\def\MN{^{\mu\nu}}
\def\mN{_\mu^\nu}

\def\R{{\mathbb R}}

\def\wh{wormhole}
\def\whs{wormholes}
\def\bh{black hole}
\def\bhs{black holes}
\def\bu{black universe}
\def\bku{black-universe}

\def\sph{spherically symmetric}
\def\ssph{static, spherically symmetric}
\def\asflat{asymptotically flat}
\def\KS{Kantowski-Sachs}

\bls{1.02}
\begin{document}
\twocolumn[
\jnumber{{\bf 17} (2), 176--180}{(2011)}

\Title{Black universes with trapped ghosts\foom 1}

\Aunames {K. A. Bronnikov\auth{a,b,2} and E. V. Donskoy\auth{b}
      }

\Addresses{
\addr a {Center of Gravitation and Fundamental Metrology,
         VNIIMS, Ozyornaya St. 46, Moscow 119361, Russia}
\addr b {Institute of Gravitation and Cosmology,
         PFUR, Miklukho-Maklaya St. 6, Moscow 117198, Russia  }
      }

%%\Rec{December 1, 2010}

\Abstract
 {A black universe is a nonsingular black hole where, beyond the horizon,
  there is an expanding, asymptotically isotropic universe. Such models
  have been previously found as solutions of general relativity with a
  phantom scalar field as a source of gravity and, without phantoms,
  in a brane world of RS2 type. Here we construct examples of static,
  spherically symmetric black-universe solutions in general relativity with
  a minimally coupled scalar field $\phi$ whose kinetic energy is negative
  in a restricted strong-field region of space-time and positive outside it.
  Thus in such configurations a ``ghost'' is trapped in a small part of
  space, which may in principle explain why no ghosts are observed under
  usual conditions.  }

\bigskip

%%%\PACS{04.20.-q, 04.20.Jb, 04.40.-b}

] %%%%%%%%%%%%%%%%%%%%%%%%%%%%%%%%%%%%%%%%%%%%%%%%%%%%%%
\Talk
\email 2 {kb20@yandex.ru}
%%\email 3 {}

\section{Introduction}

  The existence of singularities is an undesired but probably inevitable
  feature of the classical theories of gravity. Singularities are places
  where general relativity or another classical theory of gravity does not
  work, so that the theory itself reveals the limits of its validity.
  Thus a full understanding of the physics of phenomena under study (origin
  and fate of our Universe, gravitational collapse etc.) requires either
  avoidance of singularities or/and modification of the corresponding
  classical theory, or addressing quantum effects. There have been numerous
  attempts on this trend, some of them suggesting that singularities inside
  the event horizons of \bhs\ should be replaced with a kind of regular core
  ([1], see [2] for a recent review), others describing bouncing or
  ``emergent'' universes (see, e.g., [3, 4] for reviews).

  In our view, of particular interest are models that combine avoidance
  of singularities in both \bhs\ and cosmology, those which have been termed
  {\it black universes\/} [5, 6]. These are regular \bhs\ (\sph\ ones in the
  known examples) where a possible explorer, after crossing the event
  horizon, gets into an expanding universe instead of a singularity. Thus
  such hypothetic configurations combine the properties of a wormhole
  (absence of a center, a regular minimum of the area function) and a \bh\
  (a Killing horizon separating R and T regions). Moreover, the \KS\
  cosmology in the T region is asymptotically isotropic and approaches a de
  Sitter mode of expansion, which makes such models potentially viable as
  models of our accelerating Universe.

  Precisely as traversable Lo\-rentz\-ian \whs, black universes as
  solutions to the equations of general relativity require ``exotic'', or
  phantom matter, i.e., matter that violates the null energy condition.
  This can be most easily shown in the case of spherical symmetry.

  Various kinds of phantom matter are discussed in cosmology as possible
  dark energy candidates. Meanwhile, macroscopic phantom matter has not yet
  been observed. There exist theoretical arguments both {\it pro et
  contra\/} phantom fields, and the latter seem somewhat stronger, see,
  e.g., a discussion in \cite{we}.

  In \cite{we} it has been shown that black-universe models can be obtained
  without invoking phantom fields in the framework of the RS2-type
  \cite{RS2} brane-world scenario, using the modified Einstein equations
  \cite{SMS99} describing gravity on the brane. In the \bku\ solutions of
  \cite{we}, the role of exotic matter in the field equations is played by
  the ``tidal'' term of geometric origin, which has no reason to respect the
  energy conditions known for physically plausible matter fields. In such a
  scenario, wormhole solutions even without ordinary matter are also known
  \cite{bwh1}.

  In this paper, we would like to discuss another opportunity \cite{trap1}
  of obtaining \wh\ and \bku\ configurations in the framework of general
  relativity, with a kind of matter which possesses phantom properties only
  in a restricted region of space, a strong-field region, whereas far away
  from it all standard energy conditions are observed. As an example of such
  matter, we consider \ssph\ configurations of a minimally coupled scalar
  field with the Lagrangian\footnote
      {We choose the metric signature ($+,-,-,-$), the units $c = \hbar =
       8\pi G=1$, and the sign of $T\mN$ such that $T^0_0$ is the energy
       density.}
\beq                                                         \label{L_s}
      L_s = \half h(\phi) g\MN \d_\mu \phi\d_\nu \phi - V(\phi),
\eeq
  where $h(\phi)$ and $V(\phi)$ are arbitrary functions. If $h(\phi)$ has a
  variable sign, it cannot be absorbed by re-definition of $\phi$ in its
  whole range. Cases of interest are those where $h >0$ (that is, the
  scalar field is canonical, with positive kinetic energy) in a weak field
  region and $h < 0$ (the scalar field is of phantom, or ghost nature) in
  some restricted region where, e.g., a \wh\ throat can be expected. In this
  sense it can be said that the ghost is trapped. A possible transition
  between $h > 0$ and $h < 0$ in cosmology was considered in \cite{rubin}.
  Examples of such ``trapped-ghost'' \whs\ have been obtained in
  \cite{trap1}.

  The paper is organized as follows. In Section 2 we present the basic
  equations and make some general observations. In Section 3 we obtain
  explicit examples of ``trapped-ghost'' \bku\ solutions using the
  inverse-problem method, and Section 4 is a brief conclusion.

%%%%%%%%%%%%%%%%%%%%%%%%%%%%%%%%%%%%%%%%%%%%%%%%%%%% sec 2
\section{Scalar fields with a variable kinetic term}

  The general static, spherically symmetric metric can be written as
\beq
    ds^2 = A(u) dt^2 - \frac{du^2}{A(u)} - r^2(u)d\Omega^2,  \label{ds2}
\eeq
  where we are using the so-called quasiglobal gauge $g_{00} g_{11} = -1$;
  $A(u)$ is called the redshift function and $r(u)$ the area function;
  $d\Omega^2 = (d\theta^2 + \sin^2\theta d\varphi^2)$ is the linear
  element on a unit sphere. The metric is only formally static: it is really
  static if $A > 0$, but it describes a \KS\ type cosmology if $A < 0$, and
  $u$ is then a temporal coordinate. In cases where $A$ changes its sign,
  regions where $A>0$ and $A<0$ are called R- and T-regions, respectively.

  Let us specify which kinds of functions $r(u)$ and $A(u)$ are required for
  the metric (\ref{ds2}) to describe a \bu.

\begin{enumerate} \itemsep 1pt
\item
     Regularity in the whole range $u \in \R$.
\item
     Asymptotic flatness as $u \to +\infty$ (without loss of generality),
     i.e., $r(u) \approx u$, $A(u)\to 1$;
\item
     A de Sitter asymptotic as $u \to -\infty$, i.e., a T region
     ($A < 0$) where $r(u)\sim |u|$, $-A(u)\sim u^2$;
\item
     A single simple horizon (i.e., a simple zero of $A(u)$) at finite $u$.
     It is an event horizon as seen from the static side, and it is the
     starting point of the cosmological evolution as seen from the T-region.
\end{enumerate}

  The existence of two asymptotic regions $r \sim |u|$ as $u \to \pm \infty$
  requires at least one regular minimum of $r(u)$ at some $u=u_0$, at which
\beq
     r = r_0 >0, \cm r' =0, \cm r''> 0,                   \label{min}
\eeq
  where the prime stands for $d/du$. (In special cases where $r''=0$ at the
  minimum, we inevitably have $r''> 0$ in its neighborhood.)

  The necessity of violating the weak and null energy conditions at such
  minima follows from the Einstein equations. Indeed, one of them reads
\beq
      2A\, r''/r = -(T^t_t - T^u_u),                   \label{01comm}
\eeq
  where $T\mN$ are components of the stress-energy tensor (SET).

  In an R-region ($A > 0$), the condition $r''>0$ implies $T^t_t -T^u_u <
  0$; in the usual notations $T^t_t = \rho$ (density) and $-T^u_u = p_r$
  (radial pressure) it is rewritten as $\rho + p_r < 0$, which manifests
  violation of the weak and null energy conditions. It is the simplest proof
  of this well-known violation near a throat of a \ssph\ \wh\
  (\cite{m-thorne}; see also \cite{book-mifi}).

  However, a minimum of $r(u)$ can occur in a T-region, and it is then not a
  throat but a bounce in the evolution of one of the \KS\ scale factors (the
  other scale factor is $[-A(u)]^{1/2}$). Since in a T-region $t$ is a
  spatial coordinate and $u$ temporal, the meaning of the SET components is
  $-T^t_t = p_t$ (pressure in the $t$ direction) and $T^u_u = \rho$;
  nevertheless, the condition $r'' > 0$ applied to (\ref{01comm}) again
  leads to $\rho + p_t < 0$, violating the energy conditions. In the
  intermediate case where a minimum of $r(u)$ coincides with a horizon
  ($A=0$), the condition $r'' > 0$ holds in its vicinity, along with all its
  consequences. Thus the energy conditions are violated near a minimum of
  $r$ in all cases.

  Let us now turn to the scalar field $\phi(u)$ with the Lagrangian
  (\ref{L_s}). In a space-time with the metric (\ref{ds2}) it has the
  SET
\bearr
     T\mN = \half h(u) A(u) \phi'(u)^2 \diag (1,\ -1,\ 1,\ 1)
\nnn  \inch
        + \delta\mN V(u).                                    \label{SET}
\ear
  The kinetic energy density is positive if $h(\phi) >0$ and negative if
  $h(\phi) < 0$, so the solutions sought for must be obtained with $h >0$ at
  large values of the spherical radius $r(u)$ and $h < 0$ at smaller radii
  $r$. It has been shown \cite{trap1} that this goal cannot be achieved for
  a massless field ($V(\phi) \equiv 0$).

  Thus we seek \bku\ configurations with a nonzero potential $V(\phi)$,
  The Einstein-scalar equations can be written as
\bear
     (A r^2 h\phi')' - \half Ar^2 h'\phi' \eql r^2 dV/d\phi,  \label{phi}
\yy
              (A'r^2)' \eql - 2r^2 V;                         \label{00}
\yy
              2 r''/r \eql - h(\phi){\phi'}^2 ;               \label{01}
\yy
         A (r^2)'' - r^2 A'' \eql 2,                          \label{02}
\yy                                                       \label{11}
      -1 + A' rr' + Ar'^2 \eql r^2 (\half h A \phi'^2 -V),
\ear
  \eq (\ref{phi}) follows from (\ref{00})--(\ref{02}), which, given
  the potential $V(\phi)$ and the kinetic function $h(\phi)$, form a
  determined set of equations for the unknowns $r(u)$, $A(u)$, $\phi(u)$.
  \eq (\ref{11}) (the ${1\choose 1}$ component of the Einstein
  equations), free from second-order derivatives, is a first integral of
  (\ref{phi})--(\ref{02}) and can be obtained from
  (\ref{00})--(\ref{02}) by excluding second-order derivatives.
  Moreover, \eq(\ref{02}) can be integrated giving
\bear
            B'(u) \equiv (A/r^2)' = 2(3m - u)/r^4,          \label{B'}
\ear
  where $B(u) \equiv A/r^2$ and $m$ is an integration constant equal to the
  Schwarzschild mass if the metric (\ref{ds2}) is \asflat\ as $u\to
  \infty$ ($r \approx u$, $A = 1 - 2m/u + o(1/u)$). If there is a flat
  asymptotic as $u\to -\infty$, the Schwarzschild mass there is equal to
  $-m$ ($r \approx |u|$, $A = 1 + 2m/|u| + o(1/u)$.

  One more observation is that if the system contains a horizon and
  $r(u) \sim |u|$ at large $|u|$ in the T-region, then $B\equiv A/r^2$ tends
  to a finite limit, which means that there is a de Sitter asymptotic.
  Indeed, under these conditions the integral of (\ref{B'}) evidently
  converges at large $|u|$, so $B$ tends to a constant. Furthermore,
  \eq (\ref{02}) can be rewritten as $r^4 B'' + 4r^3 r' B' =-2$, hence
  $B'' < 0$ where $B'=0$, so $B(u)$ cannot have a minimum (and it is this
  circumstance that restricts the possible kinds of global causal structure
  of any scalar-vacuum solutions \cite{vac1}). This means that $B(u)$ can
  only tend to a negative constant in a T-region.

  Thus, in the Einstein-scalar system (\ref{phi})--(\ref{11}), {\it
  any\/} solution with a horizon and $r \sim |u|$ as $u\to \pm\infty$ is
  asymptotically de Sitter in the T-region. It describes a \bu\ if it is,
  in addition, \asflat\ in the R-region.

\section{Black universe models with a trapped ghost}

  If one specifies the functions $V(\phi)$ and $h(\phi)$ in the Lagrangian
  (\ref{L_s}), it is, in general, very hard to solve the above equations.
  Alternatively, to find examples of solutions possessing particular
  properties, one may employ the inverse problem method, choosing some of the
  functions $r(u)$, $A(u)$ or $\phi(u)$ and then reconstructing the form of
  $V(\phi)$ and/or $h(\phi)$. We will do so, choosing a function $r(u)$ that
  can provide a \bku\ solution. Then $A(u)$ is found from (\ref{B'}) and
  $V(u)$ from (\ref{00}). The function $\phi(u)$ is found from (\ref{01})
  provided $h(\phi)$ is known; however, using the scalar field
  parametrization freedom, we can, vice versa, choose a monotonic function
  $\phi(u)$ (which will yield an unambiguous function $V(\phi)$) and find
  $h(u)$ from \eq (\ref{01}).
\begin{figure}[ht]                     %% fig1
\centerline{\includegraphics[width=8cm]{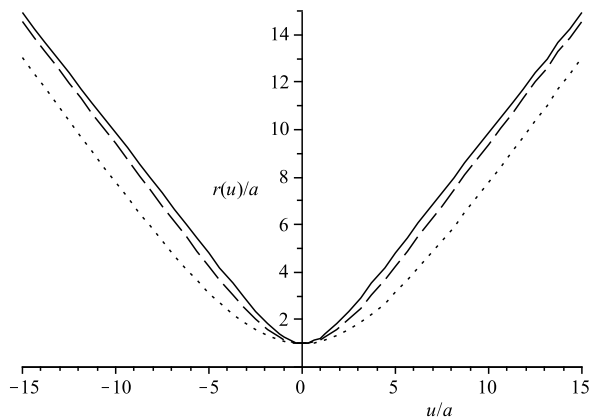}}
\caption{\small Plots of $r(u)/a$ given by \eq (\ref{r}) with $n=3,\ 5,\ 10$
    (solid, dashed, and dotted lines, respectively).
\label{fig-r}}
\end{figure}

  A simple example of the function $r(u)$ satisfying the requirements
  1--3 is (see Fig.\,1):
\beq                                \label{r}
         r(u) = a \frac{(x^2+n)} {\sqrt{x^2+n^2}},\cm
            n = \const > 2.
\eeq
  where $x = u/a$, and $a > 0$ is an arbitrary constant (the minimum radius).
  It is the same function that led to symmetric \wh\ solutions in
  \cite{trap1} under the additional assumption $m = 0$.

  Let us now integrate \eq (\ref{B'}), assuming $m > 0$. We find (see
  Fig.\,2)
\bearr\label{B}
      B(u) = \frac{3x^4 + 3x^2 n(n+1) + n^2 (n^2 + n + 1)}{3a^2(x^2 + n)^3}
\nnn \ \ \
      + \frac{mx}{8a^2 n (x^2+n)^3} \biggl[ 3x^4 (5n^2+ 2n +1)
\nnn \ \ \
      + 8 n x^2 (5n^2 + 2n -1) + 3n^2 (11n^2 -2n -1)\biggr]
\nnn \ \ \
    - \frac{3m}{8a^2 n^{3/2}} (5n^2 + 2n +1)
        \cot^{-1} \biggl(\frac{x}{\sqrt{n}}\biggr).
\ear
  The emerging integration constant is excluded by the requirement $B\to 0$
  as $u\to\infty$, providing asymptotic flatness. Examples of the behavior
  of $B(u)$ for $m=0.2a$ and some values of the parameter $n$ are presented in
  Fig.\,2.

\begin{figure}                      %% fig2
\centerline{\includegraphics[width=8cm]{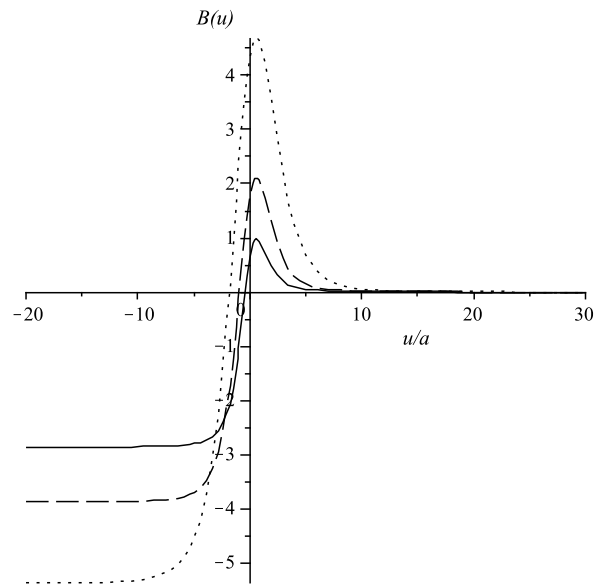}}
\caption{\small Plots of $B(u)$ given by \eq \ref{B} for $m=0.2a$, $n=3,\
	 5,\ 10$ (solid, dashed, and dotted lines, respectively).
	 \label{fig-A}}
\end{figure}

  Substituting the expressions \ref{r} and \ref{B} into \ref{00},
  taking into account that $A(u) = B/r^2$, we obtain the potential $V$ as a
  function of $u$ or $x = u/a$. This expression is rather bulky and will not
  be presented here.

\begin{figure}   %% fig3
\centerline{\includegraphics[width=8cm]{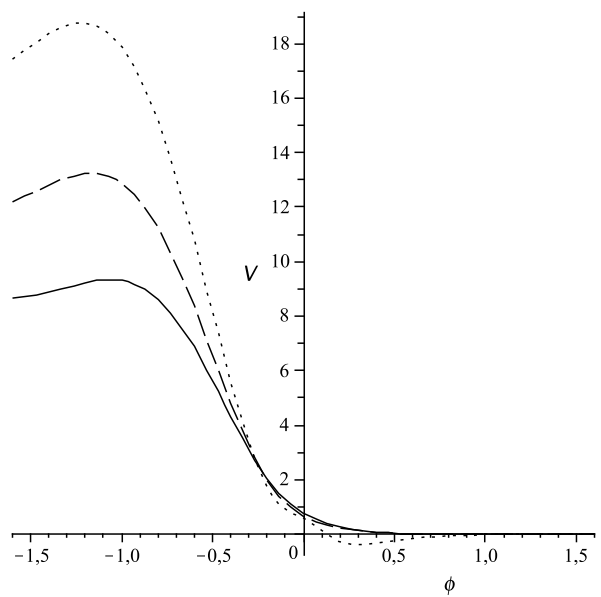}}
\caption{\small Plots of $V(u)$ for $m=0.2a$ and $n=3,\ 5,\ 10$ (solid,
	dashed, and dotted lines, respectively).  \label{figV}}
\end{figure}

  To construct $V$ as an unambiguous function of $\phi$ and to find
  $h(\phi)$, it makes sense to choose a monotonic function $\phi(u)$.
  It is convenient to assume
\beq\label{phi_2} \nq
    \phi(u) = \frac{2\phi_0}{\pi}\arctan\frac{x}{n},
    \ \ \
    \phi_0 = \frac{\pi a}{2}\sqrt{\frac{2(n-2)}{n}},
\eeq
  and $\phi$ has a finite range: $\phi \in (- \phi_0, \phi_0)$, which is
  common to kink configurations. Thus we have $x = u/a =
  n\tan(\pi\phi/2\phi_0) $, whose substitution into the expression for
  $V(u)$ gives $V(\phi)$ defined in this finite range. The function
  $V(\phi)$ can be extended to the whole real axis, $\phi \in \R$, by
  supposing $V(\phi)\equiv 0$ at $\phi \geq \phi_0$ and
  $V(\phi) = V(-\phi_0) > 0$ at $\phi < -\phi_0$.
  Plots of $V(\phi)$ are presented in Fig.\,3 for the same
  values of the parameters as in Fig.\,2.

  The expression for $h(\phi)$ is found from (\ref{01}) as follows:
\beq\label{h2}
    h(\phi) = \frac{(n-2)x^2+n^2(1-2n)} {a^2(n-2)(x^2+n)},
\eeq
  where $x = n\tan(\pi\phi/2\phi_0)$. The function $h(\phi)$ given
  by \eq (\ref{h2}) is also defined in the interval $(-\phi_0,\phi_0)$ and
  can be extended to $\R$ by supposing $h(\phi)\equiv 1$ at
  $|\phi|\geq \phi_0$. The extended kinetic coupling function
  $h(\phi)$ is plotted in Fig.\,4. Evidently, the null energy condition is
  violated only where $h(\phi) < 0$.

\begin{figure}                    %% fig4
\centerline{\includegraphics[width=8cm]{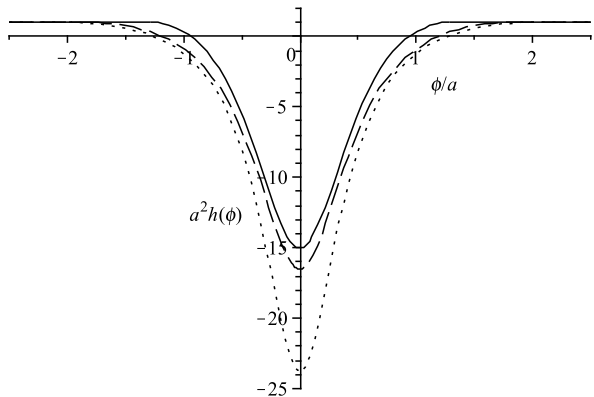}}
\caption{\small Plots of  $h(\phi)$ given by \eq \ref{h2} with $n=3,\ 5,\ 10$
        (solid, dashed, and dotted lines, respectively). \label{figh}}
\end{figure}

\section {Conclusion}

  It has been shown \cite{trap1} that a minimally coupled scalar field may
  change its nature from canonical to ghost in a smooth way without creating
  any space-time singularities. This feature, in particular, allows for
  construction of \wh\ models (trapped-ghost \whs) where the ghost is
  present in some restricted region around the throat (of arbitrary size)
  whereas in the weak-field region far from it the scalar has usual
  canonical properties. The same model has been modified here to construct
  another interesting type of configurations, black universes.

  It has also been found that, in the Einstein-scalar field system under
  study, a \ssph\ configuration is inevitably a \bu\ if it is \asflat, has a
  horizon, and the function $r(u)$ grows linearly as $u\to \pm \infty$.
  Though, all this can only happen if the scalar is of ghost nature at least
  in some part of space.

\subsection*{Acknowledgments}

    This work was supported in part by the Russian Foundation for
    Basic Research grant 09-02-00677a and by NPK MU grant at PFUR.


\small
\begin{thebibliography}{99}

\bibitem{dym92} %1
    I. Dymnikova, {\it Gen. Rel. Grav.\/} {\bf 24}, 235 (1992).

\bibitem{dym06} %2
        I. Dymnikova and E. Galaktionov,
        {\it Phys. Lett. B\/} {\bf 645}, 358 (2007).

\bibitem{bounce} %3
        P. Dzierzak, J. Jezierski, P. Malkiewicz, and W. Piechocki,
        %% Conceptual issues concerning the Big Bounce
        arXiv: 0810.3172.

\bibitem{emergent} %4
        D.J. Mulryse, R. Tavakol, J.E. Lidsey and G.F.R. Ellis,
        Phys. Rev. D {\bf 71}, 123512 (2005);\\
        S. Mukherejee, B.C. Paul, S.D. Maharaj and A. Beesham, gr-qc/0505103.

\bibitem{bron-pha1}     %5
        K.A. Bronnikov and J.C. Fabris, \PRL {96} 251101 (2006).

\bibitem{bron-pha4}     %6
        K.A. Bronnikov, V.N. Melnikov and H. Dehnen, \GRG {39} 973 (2007).

\bibitem{we}        %7
    K.A. Bronnikov and E.V. Donskoy, \GC {16} 42 (2010).

\bibitem{RS2}      %8
        L. Randall and R. Sundrum, \PRL {83} 4690 (1999), hep-ph/9906064.

\bibitem{SMS99}    %9
        T. Shiromizu, K. Maeda and M. Sasaki, \PRD {62} 024012 (2000).

\bibitem{bwh1}     %10
        K.A. Bronnikov and S.-W. Kim,
    %% ``Possible wormholes in a \bw'',
        \PRD {67} 064027 (2003), gr-qc/0212112.

\bibitem{trap1}
    K.A. Bronnikov and S.V. Sushkov, \CQG {27} 095022 (2010).

\bibitem{rubin}
    H. Kroger, G. Melkonian and S.G. Rubin,
        \GRG {36} 1649 (2004); astro-ph/0310182.

\bibitem{m-thorne}
    M.S. Morris and K.S. Thorne, Am. J. Phys. {\bf 56}, 395 (1988).

\bibitem{book-mifi}
    K.A. Bronnikov and S.G. Rubin, {\it Lectures on Gravitation and
    Cosmology\/} (MIFI press, Moscow, 2008, in Russian).

\bibitem{vac1}
    K.A. Bronnikov, \PRD {64} 064013 (2001), gr-qc/0104092;\\
    K.A. Bronnikov, \JMP {43} 6096 (2002), gr-qc/0204001;\\
    K.A. Bronnikov and G.N. Shikin, \GC {8} 107 (2002).

\end{thebibliography}
\end{document}